\documentclass{article}
\usepackage{spconf,amsmath,graphicx,caption}
\usepackage{hyperref}
\usepackage{multicol}
\usepackage{float}
\usepackage{booktabs}

\title{Attentional Speech Recognition Models Misbehave on Out-of-domain Utterances}
\name{Phillip Keung, Wei Niu, Yichao Lu, Julian Salazar, Vikas Bhardwaj}
\address{Amazon Inc.\\ \texttt{\normalsize\{keung,niuwei,yichaolu,julsal,vikab\}@amazon.com}}
\begin{document}
	\ninept
	\maketitle

	\begin{abstract}
		
		We discuss the problem of \emph{echographic transcription} in autoregressive sequence-to-sequence attentional architectures for automatic speech recognition, where a model produces very long sequences of repetitive outputs when presented with out-of-domain utterances. We decode audio from the British National Corpus with an attentional encoder-decoder model trained solely on the LibriSpeech corpus. We observe that there are many 5-second recordings that produce more than 500 characters of decoding output (i.e.\ more than 100 characters per second). A frame-synchronous hybrid (DNN-HMM) model trained on the same data does not produce these unusually long transcripts. These decoding issues are reproducible in a speech transformer model from ESPnet, and to a lesser extent in a self-attention CTC model, suggesting that these issues are intrinsic to the use of the attention mechanism. We create a separate length prediction model to predict the correct number of wordpieces in the output, which allows us to identify and truncate problematic decoding results without increasing word error rates on the LibriSpeech task.
		
	\end{abstract}
	
	\section{Introduction}
	
	End-to-end autoregressive sequence-to-sequence (AR-S2S) models \cite{attention} have become the state-of-the-art in machine translation \cite{transformer}. Since the problem of speech recognition also fits within the AR-S2S framework, these models have been successfully extended to automatic speech recognition (ASR) as well \cite{las}. Recent works show that AR-S2S models can achieve impressive word error rates on ASR tasks like LibriSpeech, Switchboard and Google Voice Search \cite{e2e_LibriSpeech,specaugment,google_e2e}. In both low-resource and high-resource settings, AR-S2S models are competitive with the various flavors of frame-synchronous hybrid (DNN-HMM) and connectionist temporal classification (CTC) based systems.
	
	However, AR-S2S ASR models were originally designed for machine translation, where the attention mechanism of the decoder can attend over the entire length of the encoded input to generate each output token; there is no notion of time-monotonicity built into the model. We investigated whether this limitation would lead to unusual behavior during decoding, and discovered the problem of \emph{echographic}\footnote{\emph{Echographia} refers to the automatic and unconscious pathological writing of text in response to external (auditory) stimuli.}\,\emph{transcription}: when AR-S2S models are used to decode out-of-domain audio, the output transcript contains the same words or phrases repeated over and over again. This behavior occurs even when the AR-S2S model performs well on the in-domain task. In other words, we found that these models are not robust to realistic variations in the input signal, leading to serious decoding issues as seen in Table \ref{decoded_ex}:
	
	\begin{table}[h]
		\caption{Decoding output from a problematic example (recording 021A-C0897X0276XX-ABZZP0, 2638.3 to 2651.6 seconds) from the BNC corpus with our AR-S2S model and a Kaldi DNN-HMM model, where both models were trained on LibriSpeech only.} 
		\label{decoded_ex}
		\centering
		\begin{tabular}{lp{5.5cm}}
			\toprule
			& Transcript \\
			\midrule
			Reference & REPRESENTATION WHAT IT IS ANYWAY PROPOSAL REPRESENTATION \\ 
			\midrule
			AR-S2S Model & FISHIN HM HM HU HU HU HU HU HU HU HU HU HU HU HU HU HU HU HU HU HU HU HU HU HU HU HU HU HU HU HU HU HU HU HU HU HU HU HU HU HU HU HU HU HU HU HU HU HU HU HU HU HU HU HU HU HU HU HU HU HU HU HU HU HU HU HU HU HU HU HU HU HU HU H \\
			\midrule
			DNN-HMM & AY AY SIR R E L M PROPOSAL REVISION \\
			\bottomrule
		\end{tabular}
	\end{table}
	
	In the sections that follow,
	
	\begin{itemize}
		\item We demonstrate that an AR-S2S ASR model which achieves reasonable word error rates (WERs) on the LibriSpeech test sets can nonetheless produce extremely long transcripts on utterances from the British National Corpus (BNC). We find that this behavior is reproducible in pretrained LibriSpeech models provided by other research groups.
		\item We identify some factors that contribute to this anomalous decoding behavior, namely the length normalization term and the attention mechanism itself.
		\item We create a separate length prediction model to decide when to truncate decoding outputs, which catches almost all of the degenerate outputs without increasing the LibriSpeech test-clean WER.
		\item We show that when the outputs are very repetitive, the attention mechanism of the AR-S2S model attends to the same section of audio without proceeding forwards in time. This may explain why we do not see echographic transcriptions in a standard frame-synchronous hybrid model.
		
	\end{itemize}
	
	\section{Materials and methods}
	
	\subsection{Datasets}
	
	\begin{figure*}[h]
		\centering
		\begin{multicols}{2}
			\includegraphics[width=6.5cm,trim={0 0.8cm 0 3.4cm},clip]{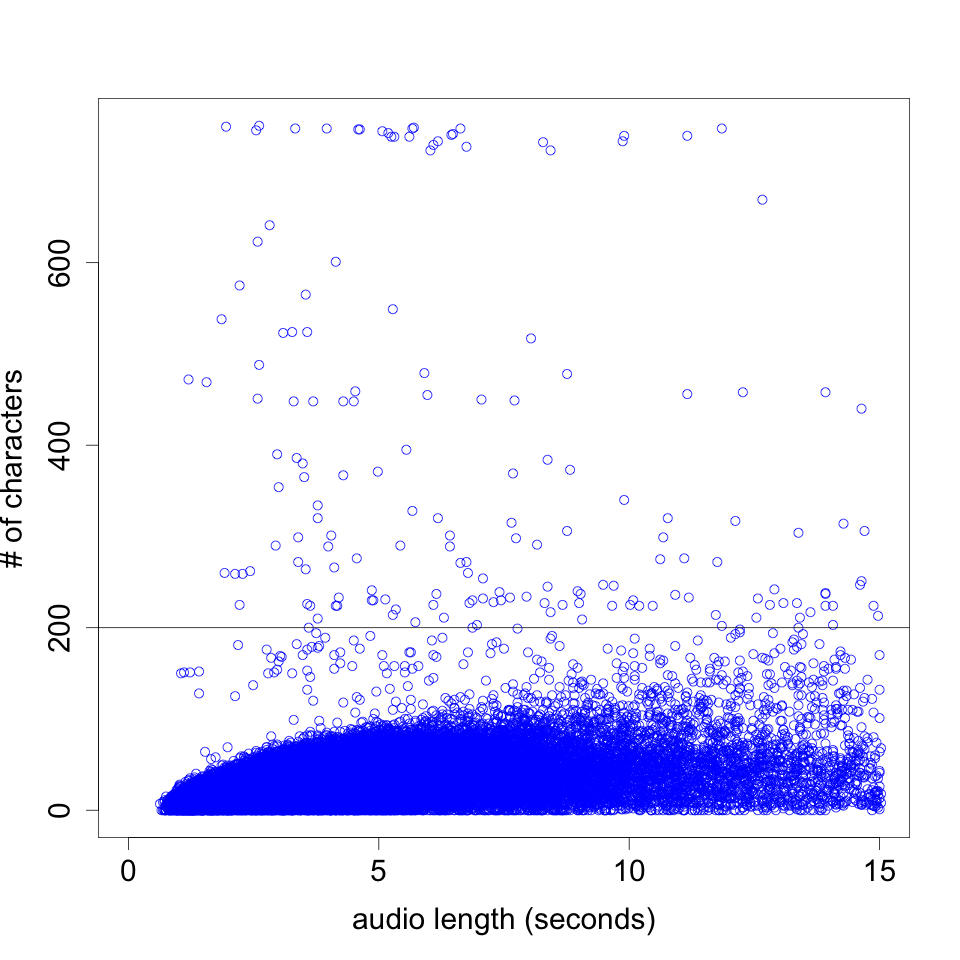} \par
			\includegraphics[width=6.5cm,trim={0 0.8cm 0 3.4cm},clip]{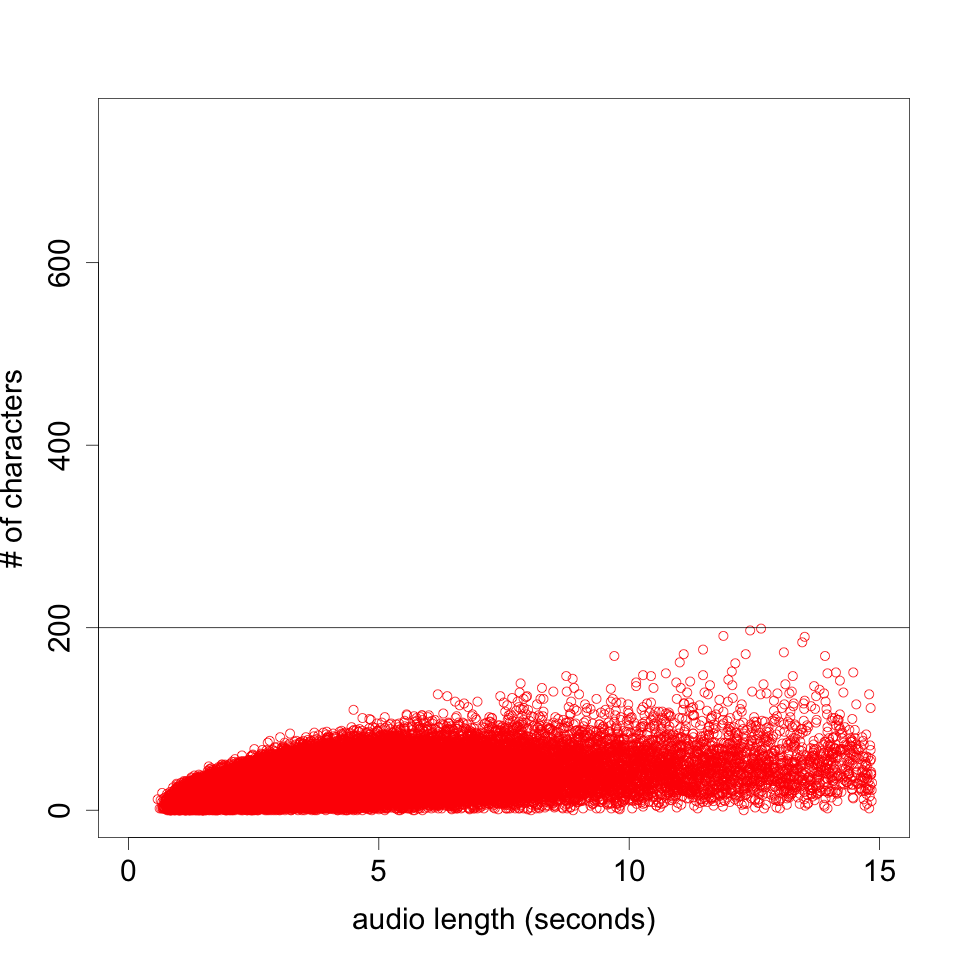}
		\end{multicols}
		
		\caption{Number of characters in the model's output versus the length of the audio (seconds) on 39,323 utterances from the BNC corpus. The line at 200 characters is our cutoff for `echographic' utterances. (There are 170 such utterances with AR-S2S decoding, and 0 with DNN-HMM decoding.) With the AR-S2S ASR model (left panel), there are outputs which are over 700 characters long, but such outliers do not exist for the frame-synchronous DNN-HMM model (right panel).}
		\label{bad_char_v_seconds}
	\end{figure*}
	
	Our ASR models were trained on the 960h LibriSpeech corpus \cite{LibriSpeech}, which is primarily composed of audio from US English speakers reading passages out of books in the public domain. We collected the out-of-domain recordings from the British National Corpus (BNC) \cite{bnc}, which contains spontaneous conversations from UK English speakers collected in the early 1990s with portable tape recorders and significant background noise. Using the provided word-level alignments (found at \url{http://www.phon.ox.ac.uk/AudioBNC}), we selected 39,323 5 to 15-word utterances at random across the recordings in BNC. The utterance lengths were capped at 15 seconds. 
	
	From the set of 39,323 utterances, we identified 170 `echographic' examples, which we define as utterances that produce at least 200 characters of output upon decoding with our AR-S2S model. Some AR-S2S model transcriptions are extremely long, exceeding 600 characters in length. Upon manual review of the data, we found that the transcripts based on the provided word-level alignments were often incorrect. Therefore, we also created a \textit{clean} evaluation subset of 1,607 utterances, which were selected from utterances where the AR-S2S decoding result had an utterance-level WER below 50\%.
	
	The data (along with download URLs) needed to recreate our BNC datasets can be found at \url{https://github.com/aws-samples/seq2seq-asr-misbehaves} .
	
	\subsection{Models}
	\label{model_details}

	Our AR-S2S ASR model is an attentional encoder-decoder model, with a 6-layer bidirectional LSTM encoder and a 2-layer transformer decoder, implemented in MXNet with components from GluonNLP \cite{gluonnlp}. We chose the model hyperparameters by tuning for the best WER on LibriSpeech dev-clean. The encoder and decoder layers have 384 hidden units. In a manner similar to \cite{frame_stacking}, frames of 64-dimensional log-filterbank energy (LFBE) features are stacked together for an effective frame rate of 30 ms. The output vocabulary is 10,025 subwords generated through byte pair encoding (BPE) \cite{bpe}. We applied a label smoothing of 0.05 \cite{label_smoothing} to the output probabilities. We used the Adam optimizer and a newbob learning rate schedule for model training. We also trained a 2-layer 2048-unit LSTM language model (LM) on the LibriSpeech text corpus. We performed decoding jointly with the LM via shallow fusion \cite{asr_lm, shallow_fusion}. Layer normalization \cite{layernorm} is applied between layers of the encoder. We apply length normalization \cite{length_penalty} with $K=5$ and $\alpha=1.0$. The WER of our AR-S2S ASR model is 4.7\% on test-clean and 13.5\% on test-other of the LibriSpeech corpus.
	
	The DNN-HMM model is a Kaldi model trained on the LibriSpeech corpus. It (and its decoding recipes) are publicly available for download at \url{http://kaldi-asr.org/downloads/build/10/trunk/egs/LibriSpeech/s5/} . The WER of the DNN-HMM model in \cite{LibriSpeech} is 5.5\% on LibriSpeech test-clean and 13.9\% on test-other.
	
	\section{Results}
	
	\subsection{Example of unexpected decoding behavior}
	
	In Table \ref{decoded_ex}, we show the AR-S2S model's output on a problematic BNC example. The utterance is 13.3 seconds long, and the AR-S2S model produced 230 characters of output (i.e.\ 17.3 characters per second). We see that the model output contains many instances of `HM' and `HU', neither of which occur in the audio.
	
	\subsection{Reproducibility across architectures and implementations}
	
	We confirmed that this issue is reproducible in different settings and frameworks:
	
	We considered the self-attention CTC (SAN-CTC) model in \cite{san}, which is a character-level model trained with the CTC loss and a self-attention encoder architecture. We use the same model as in the paper, with a WER of 4.8\% on test-clean and 13.1\% on test-other of the LibriSpeech corpus. Decoding is done via a WFST compiled from the included 4-gram LM \cite{eesen}, with a weight of 0.48 on acoustic log-probabilities, a CTC blank scale of 0.3, and a decoding beam size of 17. We found that 34 out of 39,323 BNC utterances still produce at least 200 characters, with some of those utterances being less than 10 seconds long.
	
	We also considered an independent implementation of an AR-S2S model from ESPnet \cite{espnet}, where we used the model artifact provided at \url{https://drive.google.com/open?id=1BtQvAnsFvVi-dp_qsaFP7n4A_5cwnlR6} . This transformer-based model \cite{speech_trans} was trained on the LibriSpeech corpus, and incorporated an auxiliary CTC loss in addition to the standard AR training loss. We used the default decoding parameters except for the beam width, which we reduced to 8 to speed up the decoding process. Under this setting, the model achieves a WER of 2.7\% on LibriSpeech test-clean and 5.9\% on test-other. We found that 95 out of 39,323 BNC utterances produce at least 200 characters of output. We continue to observe the repetitive decoding issues, despite using a near state-of-the-art AR-S2S model that includes a CTC loss term.
	
	\subsection{Number of characters produced per second of audio}
	
	\begin{figure*}[h]
		\centering
		\begin{multicols}{2}
			\includegraphics[width=6cm]{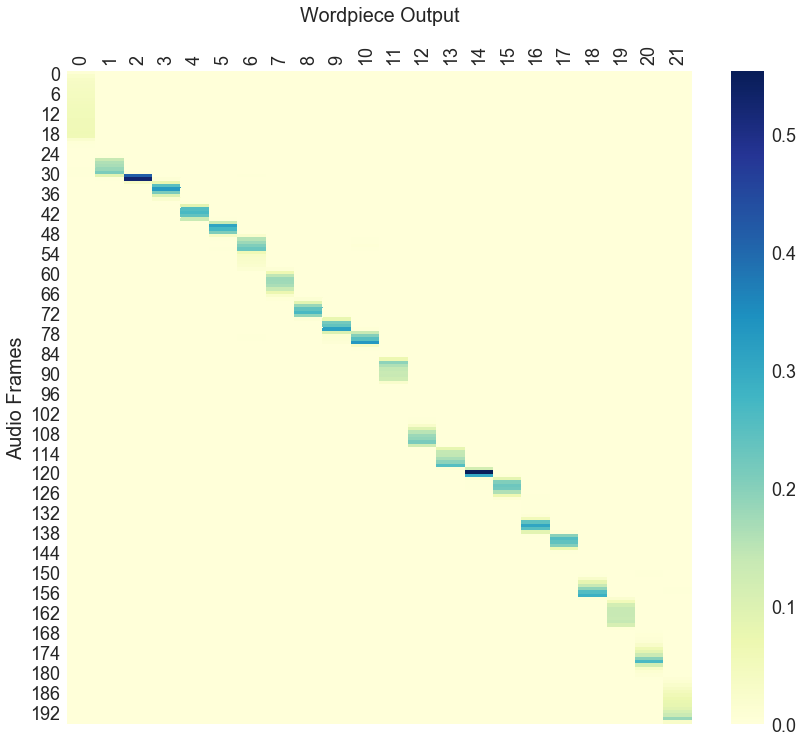} \par
			\includegraphics[width=6cm]{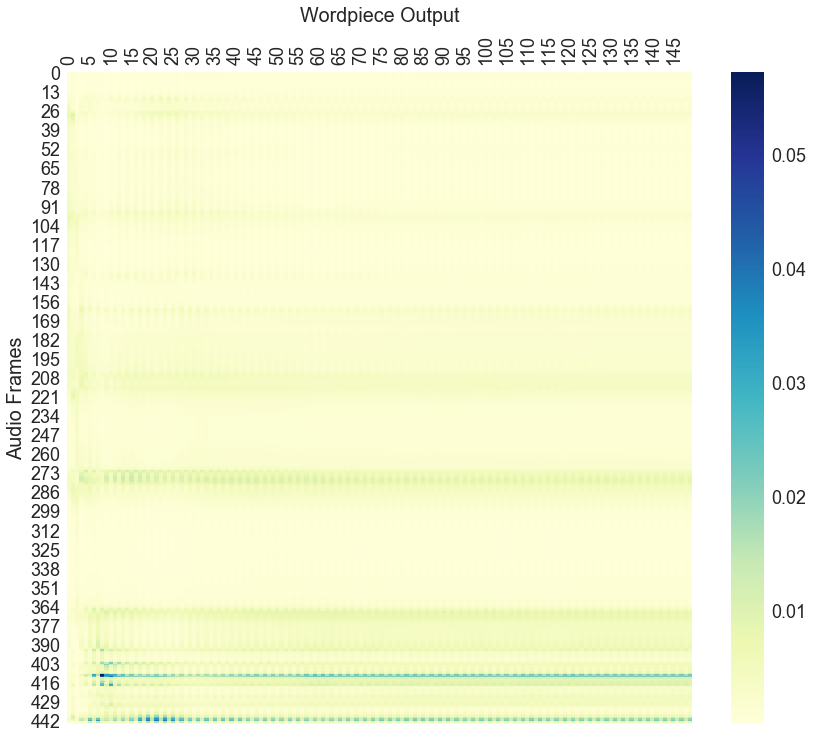}
		\end{multicols}
		
		\caption{Decoder attention weights on the audio from the AR-S2S model. The vertical indices correspond to audio frames, and the horizontal indices to the wordpiece outputs. On the left, we decode an example from LibriSpeech dev-clean. Here, the attention mechanism proceeds monotonically as expected. On the right, we decode the problematic example from Table \ref{decoded_ex}.  We see that the attention mechanism does not proceed monotonically along the frames, and gets stuck near the end of the audio.}
		\label{att_weights}
	\end{figure*}
	In Figure \ref{bad_char_v_seconds}, we plot the length of the output (in characters) against the length of the audio (in seconds). Generally speaking, we expect the length of the output to increase linearly with the number of seconds in the recording.
	
	The AR-S2S model does show a roughly linear relationship between the length of the audio in seconds and the number of characters, but there are many outliers present in the figure. In particular, there are utterances which induce the model to produce over 100 characters per second, which is unexpected behavior. There is a ceiling at about 750 characters caused by the maximum number of wordpieces produced during beam search decoding, which is capped at 150.
	
	The DNN-HMM model shows the expected relationship between the length of the audio in seconds and the number of characters. There are no obvious outliers present in the figure, even on the examples where the AR-S2S model misbehaves.
	
	\subsection{Behavior of attention weights}
	
	In Figure \ref{att_weights}, we plot the attention weights of the top layer of the decoder as it decodes audio. On the left, we decode an example from LibriSpeech dev-clean. Here, the attention mechanism proceeds monotonically as expected. On the right, we decode the problematic example from Table \ref{decoded_ex}. We can see that the attention mechanism does not proceed monotonically along the frames. In addition, the decoder's attention repeatedly passes over the same region of the encoded audio for many time steps. 
	
	The attention mechanism's behavior suggests that enforcing strict monotonicity in the decoder may alleviate the problem of overly long decoding outputs, though at the cost of potentially degrading performance on all utterances. Monotonic attention for speech was explored in \cite{online_att} and is beyond the scope of this paper.
	
	\subsection{Effect of decoder settings}
	\label{section_LN}
	
	We evaluate whether different decoder parameters are to blame for the problem of echographic transcription. Surprisingly, changing the beam width across 1, 2, 4, 8 and 16 and the LM weight across 0.0, 0.125, 0.25 and 0.375 had no effect on the repetitive decoding outputs, and we do not present those results here.
	
	We found that machine translation-style length normalization \cite{length_penalty} did have a significant impact on the number of problematic outputs remaining, at the cost of accuracy. At each step during beam search, we divide the log-probability by a normalization term:
	
	\[ \text{LP}(Y) = \frac{(K + |Y|)^\alpha}{(K+1)^\alpha},  \]
	
	\noindent where $|Y|$ is the current length of the hypothesis, and $K\geq0$ and $0 \leq \alpha \leq 1$ are hyperparameters. When $\alpha$ is 0, this is the same as performing standard beam search with the log-probability alone.
	
	The length normalization term is designed to encourage the model to select longer hypotheses. The intuition behind the length normalization term is that during beam search, the decoder favors shorter outputs, since the log-probability contribution from an additional word is always negative. The normalization term compensates for this by making the scores of longer hypotheses larger (i.e.\ closer to 0, since we are working on the log-probability scale.) The larger the $\alpha$, the more the longer hypotheses are favored.
	
	In Table \ref{length_penalty}, we see that the number of echographic examples goes down as the length normalization $\alpha$ goes to 0. However, the performance on the evaluation sets also drops significantly. Even at $\alpha=0.0$, 11 out of 170 echographic examples continue to yield 200 or more characters of output.
	
	\begin{table}[th]
		\caption{Number of echographic examples remaining (out of 170) as a function of the length normalization hyperparameter $\alpha$. The other decoding hyperparameters are held fixed at the setting in Section \ref{model_details}. The in-domain evaluation set is LibriSpeech dev-clean, and the out-of-domain evaluation set is our clean BNC subset.}
		\label{length_penalty}
		\centering
		\begin{tabular}{lccc}
			\toprule
			\textbf{$\alpha$} & \textbf{Echographic Utts.} & \textbf{Libri.\ WER} & \textbf{BNC WER} \\ 
			\midrule
			1.0 & 170/170 & 4.7 & 38.3 \\ 
			0.8 & 66/170 & 4.9 & 38.6 \\ 
			0.6 & 26/170 & 5.3 & 38.8 \\ 
			0.4 & 13/170 & 5.9 & 39.6 \\ 
			0.2 & 11/170 & 6.4 & 40.1 \\ 
			0.0 & 11/170 & 6.7 & 40.6 \\ 
			\bottomrule
		\end{tabular} 
	\end{table}
	
	\subsection{Truncation via output length prediction}
	
	We address the decoding problem directly by creating a model that predicts the correct length of the output. We truncate the decoding result if its length exceeds a multiple $\eta$ of the predicted output length, where $\eta$ is a hyperparameter.
	
	We model the number of output wordpieces as a Poisson distribution parameterized by an LSTM network taking acoustic inputs. If $N$ is the number of wordpieces for the sequence of LFBE frames $X$ of length $T$, then
	\[ p(N=n|X) = \frac{\Lambda_\theta(X)^n e^{-\Lambda_\theta(X)}}{n!}  \]
	where 
	\[\Lambda_\theta(X) = \sum^T_{t=1} \text{ReLU}(\alpha + \beta^T f_\theta(X)_t) \]
	and $f_\theta$ is a bidirectional LSTM network, and $\alpha$ (scalar), $\beta$ (vector) are learned parameters for an affine transformation. The predicted output length is $\text{round}(\Lambda_\theta(X))$, which is the integer closest to the Poisson expectation $\Lambda_\theta(X)$ for the input $X$.
	
	We use the AR-S2S model's bidirectional LSTM encoder as an initialization for $f_\theta$. We finetune the LSTM model on LibriSpeech data. For each utterance, $N$ is the number of wordpieces in the reference transcript, and $X$ is the sequence of LFBE frames derived from the utterance.
	
	We used LibriSpeech dev-clean to determine the model hyperparameters. The mean absolute error of the length prediction model on LibriSpeech test-clean is 2.7 wordpieces. We truncate all decoding outputs at a multiple of the predicted output length. For example, if the multiple $\eta$ is 1.1 and the predicted number of wordpieces $\hat{N}$ is 10 for the input $X$, then only the first 11 wordpieces of the decoding output for $X$ are retained. This has no effect if the number of outputs is less than $\eta$ times the predicted length.
	
	In Table \ref{truncation}, we observe that truncating at $\eta=$ 1.3 times of the predicted length has no impact on the LibriSpeech WER, and filters out all but 10 of the 170 repetitive examples. This is in contrast to approaches like length normalization (Section \ref{section_LN}), where lower rates of echographic transcription came at the cost of increased WERs.
	\begin{table}[th]
		\caption{Number of echographic examples remaining (out of 170) as a function of the length multiple $\eta$. We truncate the decoding results at different multiples of the predicted length. The decoding hyperparameters are held fixed at the default setting in Section \ref{model_details}. The in-domain evaluation set is LibriSpeech dev-clean, and the out-of-domain evaluation set is our clean BNC subset.}
		\centering
		\label{truncation}
		\begin{tabular}{lccc}
			\toprule
			\textbf{$\eta$} & \textbf{Echographic Utts.} & \textbf{Libri.\ WER} & \textbf{BNC\ WER} \\ 
			\midrule
			1.0 & 4/170 & 11.4 & 51.2  \\ 
			1.1 & 7/170 & 7.4 & 42.9 \\ 
			1.2 & 7/170 & 5.6 & 39.1 \\ 
			1.3 & 10/170 & 4.7 & 38.4 \\ 
			\bottomrule
		\end{tabular} 
	\end{table}

	\section{Related work}
	
	Adversarial examples that induce unexpected model predictions have been studied extensively in the machine learning literature. In the context of speech processing, \cite{adv_audio} constructs adversarial audio to obtain arbitrary transcriptions, and \cite{adv_verification} generates adversarial examples to fool speaker verification models. In the context of computer vision, \cite{adv_ex} showed that small perturbations to the input image lead to very different classification results, and \cite{robust_adv} extends this idea to the case of 3D adversarial images.
	
	In the aforementioned work, the authors found input perturbations that led to unexpected outputs in models. However, the specific issue of anomalous decoding behavior on \emph{natural} audio with AR-S2S models is not well-documented. The closest work that we know of is \cite{battenberg}, which mentioned that decoding silence with an AR-S2S ASR model generates garbled text, but our own experiments with digital silence and white noise did not replicate that finding. To the best of our knowledge, no prior work has demonstrated that decoding recordings of natural speech with AR-S2S ASR models can induce pathological decoding behavior. We have found (rather than generated) natural adversarial examples \cite{natural_adv} for these models.
	
	\section{Conclusion}
	
	We showed that out-of-domain utterances induce unusual behavior in AR-S2S and self-attention CTC models at decoding time, and that this behavior does not appear with DNN-HMM systems. Changes to common decoder parameters (i.e.\ beam width and LM weight) had no meaningful effect on this behavior. Adjusting the length normalization term of the decoder had an effect, but it had a significant negative impact on the WER.
	
	We then trained a bidirectional LSTM model to predict the correct number of wordpieces for a given piece of audio, which we used to detect and truncate degenerate decoding results. This procedure removed the most egregious examples without degrading the WER on the LibriSpeech task. 
	
	It is surprising that, under the default settings that were optimized on the in-domain LibriSpeech dev sets, our AR-S2S implementation and the ESPnet implementation produced extremely long, highly repetitive outputs on more than 90 examples from the BNC corpus. Future work could explore potential changes to model architectures and/or decoder algorithms to avoid this issue altogether.
	
	\section{Acknowledgments}
	
	We would like to thank Katrin Kirchhoff and Jason Sun for the helpful discussions and for their support of our research endeavors.

\bibliographystyle{IEEEbib}
\bibliography{refs}

\end{document}